# Atom-Specific Probing of Electron Dynamics in an Atomic Adsorbate by Time-Resolved X-ray Spectroscopy

Simon Schreck[1], Elias Diesen[2]*, Martina Dell'Angela[3], Chang Liu[1], Matthew Weston[1], Flavio Capotondi[4], Hirohito Ogasawara[5], Jerry LaRue[6], Roberto Costantini[3,7], Martin Beye[8], Piter S. Miedema[8], Joakim Halldin Stenlid[1,2], Jörgen Gladh[1,5], Boyang Liu[1], Hsin-Yi Wang[1], Fivos Perakis[1], Filippo Cavalca[1], Sergey Koroidov[1], Peter Amann[1], Emanuele Pedersoli[4], Denys Naumenko[4], Ivaylo Nikolov[4], Lorenzo Raimondi[4], Frank Abild-Pedersen[2], Tony F. Heinz[5,9], Johannes Voss[2], Alan C. Luntz[2], and Anders Nilsson[1]

[1]Department of Physics, AlbaNova University Center, Stockholm University, SE-10691 Stockholm, Sweden

[2]SUNCAT Center for Interface Science and Catalysis, SLAC National Accelerator Laboratory, 2575 Sand Hill Road, Menlo Park, California 94025

[3]CNR-IOM, SS 14 – km 163.5, 34149 Basovizza, Trieste, Italy

[4]FERMI, Elettra-Sincrotrone Trieste, SS 14 – km 163.5, 34149 Basovizza, Trieste, Italy

[5]SLAC National Accelerator Laboratory, 2575 Sand Hill Road, Menlo Park, California 94025

[6]Schmid College of Science and Technology, Chapman University, Orange, California 92866

[7]Physics Department, University of Trieste, Via Valerio 2, 34127 Trieste, Italy

[8]Deutsches Elektronen-Synchrotron DESY, Notkestrasse 85, Hamburg 22607, Germany

[9]Department of Applied Physics, Stanford University, Stanford, California 94305

*Present address: Fritz-Haber-Institut der Max-Planck-Gesellschaft, Faradayweg 4-6, D-14195 Berlin, Germany. E-mail: diesen@fhi.mpg.de

## ABSTRACT

The electronic excitation occurring on adsorbates at ultrafast time scales from optical lasers that initiate surface chemical reactions is still an open question. Here, we report the ultrafast temporal evolution of X-ray absorption spectroscopy (XAS) and X-ray emission spectroscopy (XES) of a simple well known adsorbate prototype system, namely carbon (C) atoms adsorbed on a nickel (Ni(100)) surface, following intense laser optical pumping at 400 nm. We observe ultrafast (~100 fs) changes in both XAS and XES showing clear signatures of the formation of a hot electron-hole pair distribution on the adsorbate. This is followed by slower changes on a few ps time scale, shown to be consistent with thermalization of the complete C/Ni system. Density functional theory spectrum simulations support this interpretation.

## BODY TEXT

Fundamental dynamical processes of adsorbates on surfaces, *e. g.* energy or charge transfer, desorption, *etc.* often define catalytic and especially photocatalytic activity and selectivity. These processes are typically in the femtosecond regime, where optical pump-probe experiments have been used to follow the adsorbate dynamics [1–3]. The optical pump excites high energy electron-hole (e-h) pairs in the metal which locally thermalize within ~100 fs to create a quasi-equilibrium, often described by a two temperature model (2T) with one high temperature for the e-h pairs and one for substrate phonons [4,5], with equilibration between the two systems occurring on a several ps time scale. Both modes can excite vibrations of adsorbates and therefore ultimately drive chemistry. The focus of laser-induced chemistry on metals has traditionally been on the relative role of substrate vibrations or electronic excitations in inducing chemical change [6]. Inferences regarding the electronic structure changes of adsorbates due to optical laser excitation of the supporting metal have only been *indirect*, based on what surface chemistry is induced [5]. However, *direct* experimental probing of the electronic structure changes due to the laser excitation on the adsorbate atoms themselves has been lacking.

Recently, with the emergence of X-ray free-electron lasers, optical pump – X-ray probe measurements of ultrafast surface chemistry have become possible. Since the X-ray spectroscopic probe can measure element-specific changes in the electron structure, much more detailed information about the chemistry occurring in the ultrafast time regime can be obtained than in pure optical pump-probe experiments. For example, it has been possible to detect a transient precursor state in CO desorption [7,8], a transient adsorbed CHO intermediate [9], the transition state region in CO oxidation [10], and ultrafast vibrational excitations in adsorbed CO on a Ru(0001) surface [11], as well as *e.g.* elucidate photooxidation of CO on oxides [12]. However, in none of the previous work done at X-ray lasers has it been possible to detect electronic structure changes on the adsorbate prior to the nuclear dynamics. In contrast, for bulk metals direct X-ray probing of the excited electronic system has been achieved [13,14].

In this Letter we aim to provide, in an atom-specific way, this missing information on direct changes of the adsorbate electronic structure caused by an optical pulse. We choose a most simple, strongly adsorbed atomic system where the electronic structure is well understood in terms of the d-band model, which describes the simple adatom 2p interaction with the metal states leading to bonding and antibonding adatom 2p – metal d states [15]. With X-ray spectroscopy the projection on the adsorbate atom [16] of the bonding and antibonding states can be directly experimentally probed, providing an opportunity for a fundamental picture of optical laser induced changes. The carbon adsorbate is a prototypical strongly bonded atomic adsorbate that is well characterized with X-ray spectroscopy-based synchrotron radiation measurements because of its relatively narrow linewidths [17,18]. Here, we directly follow the changes in the occupied and unoccupied local states of a strongly adsorbed C atom, using X-ray absorption spectroscopy (XAS) and X-ray emission spectroscopy (XES) after intense laser excitation (at 400 nm) of a Ni(100) substrate. With the usage of externally seeded X-ray laser beams [19], it is possible to probe the adsorbate electronic structure dynamics with a time resolution of ≈ 100 fs. We observe both ultrafast (~100 fs) changes in the C X-ray spectra and slower changes occurring on a ps time-scale. Both electronic and nuclear excitations of C thus affect the X-ray spectra, and these excitations are analyzed theoretically.

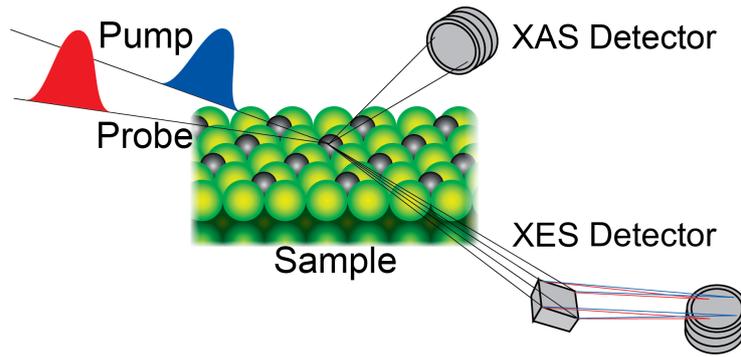

FIG. 1. A 400 nm pump laser and a soft X-ray probe laser beam are incident at grazing angle to the C/Ni(100) surface. The two beams arrive at different times allowing for pump-probe measurements. The XAS detector detects all emitted photons in a large solid angle whereas the XES detector contains an energy-dispersive soft X-ray spectrometer. In XAS the incoming photon energy is scanned whereas in XES it is fixed to allow for specific excitations.

Fig. 1 shows the principles of core-level excitation (XAS) and de-excitation (XES) allowing for element-specific probing of the electronic structure in both the occupied (XES) and unoccupied (XAS) states through the involvement of the strongly localized C 1*s* level [17]. The measured spectra reflect the projected density of states of C 2p character through the dipole selection rule for a 1s-2p transition. To obtain a temporally well-controlled probe of soft X-ray pulses relative to the optical pump, a seeded X-ray laser was used from the FERMI facility [20]. The sample is adsorbed C on Ni(100) (coverage 0.5 ML) that has been structurally characterized as a *p4g* overlayer [21] with the C atoms fourfold-coordinated by Ni atoms in a hollow site and with the C almost within the first Ni layer [22]. The 400 nm (3.1 eV photon energy) optical pump laser, with a focus of ~150×150 $\mu m^2$ and a pulse energy of ~152 µJ, arrives at the sample, followed by the soft X-ray probe at variable time delays. The C atom probing is performed with XAS using a fluorescence detector and with XES using a soft X-ray spectrometer. XAS and XES studies of C/Ni have shown relatively narrow spectral profiles allowing observation of detailed changes [23]. Details on the experimental method is found in the Supplemental Material [24].

Fig. 2(a) shows the C K-edge XAS at different delay times. The electric field (E-) vector is parallel to the surface, meaning the XAS probes unoccupied orbitals of local C 2p character oriented parallel to the surface. The spectrum at -0.4 ps delay (i.e., before the pump pulse) can be directly compared to XAS of the same system measured with synchrotron radiation, where the photon energy has been calibrated on an absolute scale [18]. This gives the energy scale in Fig. 2(a). For adsorbates on metallic systems, the onset of the XAS corresponds to the Fermi level, so that the core-level binding energy can be used to reference the XAS onset [18,25,26]. The XPS-determined C1s binding energy of C on Ni(100) is 283.0 eV and is indicated with an arrow in Fig. 2(a). The broad peak in the XAS spectrum that is above the XPS binding energy position can be related to the Ni 3d – C 2p antibonding resonance [15–17]. The weak structure around 282 eV is due to imperfect background subtraction.

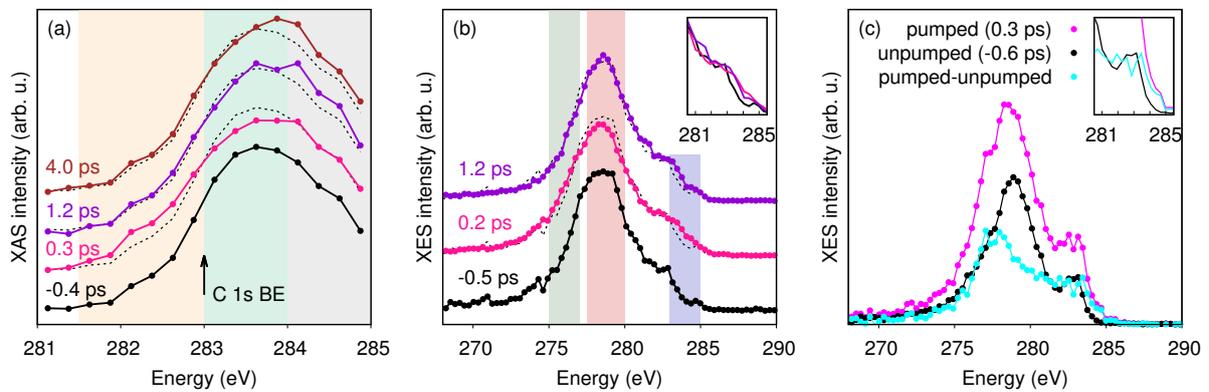

FIG. 2. Experimental XAS and XES spectroscopy of C/Ni(100). (a) XAS spectra measured with the E-vector parallel to the surface at different delay times between the optical and soft X-ray laser beams. The arrow indicates the C 1s binding energy used for calibration. (b) XES spectra in grazing emission using an excitation energy of 284 eV. (c) Resonant excited XES spectra at a photon energy of 282.5 eV with negative delay and 0.3 ps delay, and the difference spectrum between the two. The inserts in (b) and (c) show an enlarged region around the C1s core-level binding energy corresponding to the Fermi level of the system. The color coding in (a) and (b) indicates regions where spectral intensity variations with delay time are shown in fig. 3.

The XAS spectrum at 0.3 ps delay shows two significant changes: increased intensity below 283 eV (orange region in Fig. 2(a)), and reduced intensity above 283 eV (turquoise region). The use of the C K-edge XAS directly implies that the observed changes occur in the electronic structure on the adsorbed C atoms. These two changes are diminished in the spectrum at 1.2 ps delay and instead additional intensity is seen in the region 284-285 eV (grey region). At 4.0 ps delay the spectrum is nearly identical to that at 1.2 ps.

Fig. 2(b) shows C K-edge XES spectra generated with an incoming photon energy of 284 eV with the E-vector parallel to the surface and the emitted soft X-rays detected at grazing angle with respect to the surface. With grazing emission we probe the occupied C 2p-derived orbitals both parallel and perpendicular to the surface [17]. The emission energy scale for the spectrum at negative delay is calibrated by comparison with XES from synchrotron radiation [17]. At -0.5 ps delay the spectrum consists of a broad main peak centered around 278.5 eV with a clear asymmetry to higher energies and a cut-off at 283 eV corresponding to the Fermi level. The main peak is associated with the Ni 3d – C 2p bonding resonance. After laser excitation at 0.2 ps delay we observe 3 different changes in the XES spectrum, i) the main peak at 278.5 eV loses some intensity (red region), ii) there is small additional intensity appearing between 275-277 eV (green region) and iii) additional intensity above the Fermi level at 283 eV (blue region). These changes vanish in the 1.2 ps delay spectrum, except for a small intensity increase still remaining above 283 eV. It is important to note that the changes around the main peak (i) and below (ii) correspond to electronic states that are too far below the Fermi level to be energetically accessible for a single-photon excitation from the 400 nm (3.1 eV) laser light.

Fig. 2(c) shows XES spectra at 0.3 ps delay and for the unpumped case (-0.6 ps delay) after X-ray excitation to 282.5 eV, *i. e.* 0.5 eV below the Fermi edge. In the unpumped spectrum, the intensity in this (nominally Pauli blocked) region is due to phonon, core-hole lifetime, and instrumental broadening of the XAS spectrum, giving rise to a low-energy tail [18]. Both spectra are normalized by the incoming fluence; the total intensity therefore increases after pumping, corresponding to the increased XAS intensity at 282.5 eV at 0.3 ps delay. The difference spectrum between the pumped and unpumped corresponds to selective probing of atoms where holes have been generated below the Fermi level by the 400 nm laser excitation increasing the overall X-ray absorption cross section and thereby the total XES intensity. We observe in the difference XES spectrum that the main peak is shifted from 279 eV to 277.5 eV, and additional intensity appears above the Fermi level. This is consistent with the XES spectral changes due to the laser pump shown in Fig. 2(b), but more enhanced since the selective X-ray excitation leads to a decreased background signal from non-excited species.

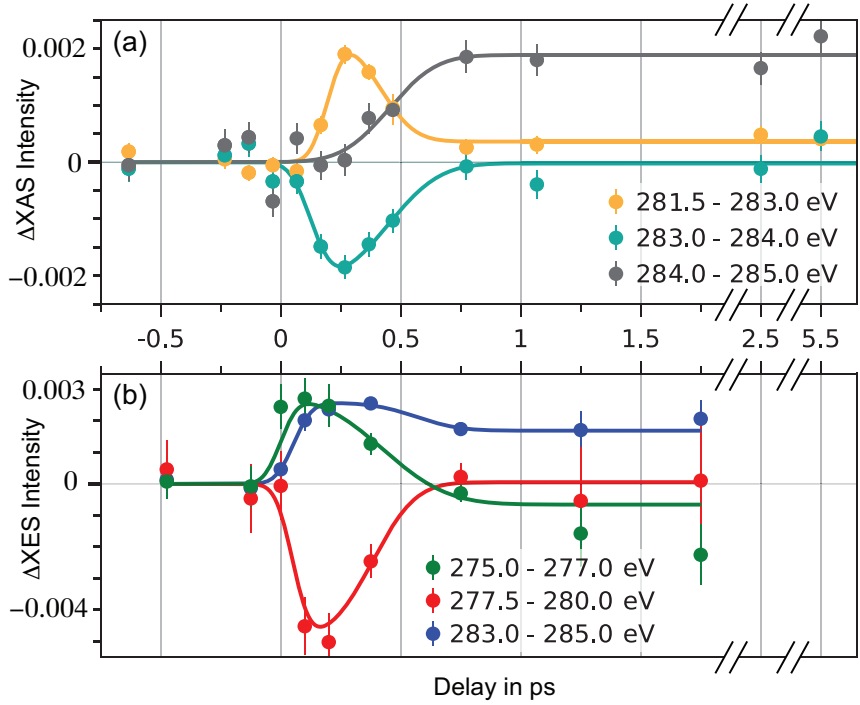

FIG. 3. XAS and XES spectral regional integrated difference intensity of different delay times. (a) XAS time traces and (b) XES time traces (excitation at 284 eV). The color of each time trace represents different spectral regions indicated in Fig. 2. Note that these regions are not the same for XAS and XES.

Fig. 3 shows the time traces from the spectral changes in Fig. 2(a) and (b). Starting with the XAS time-dependent changes shown in Fig. 3(a), we observe changes for both the holes (orange) and electrons (turquoise) at 0.25 ps with an apparent Gaussian rise time of around 0.1 ps, and with a slower recovery of around 0.2 ps. There is a delayed onset of around 0.15 ps as determined by the Gaussian rise half maximum. The long-term spectral change (grey) is delayed until 0.5 ps with a rise time of 0.2 ps with no detectable recovery within the measured 4.0 ps. The XES spectral changes around the main peak (green and red) appear to be faster with the onset at almost 0 ps and with a slightly faster rise time; the recovery time of ~ 0.25 ps appears somewhat slower than in the XAS. Both the XAS and XES rapid changes have decayed within 0.5 ps. The changes above the Fermi level (blue) in the XES spectra have an onset of 0.1 ps and then remain, similar to the long-term time component in the XAS. We note that the XAS and XES data were not measured simultaneously and it is not possible to exclude timing drift defining the temporal overlap of the optical and X-ray beam ($t_0$) over 24 hrs, so that uncertainties in defining $t_0$ ~ ±0.1 ps are realistic.

In order to assist in the interpretation of the experimental data we have computed X-ray spectra with both electronic and nuclear excitations. This was done by density functional theory (DFT) calculations at the GGA level using the Quantum ESPRESSO [27,28] code, with spectra calculated using the xspectra [29–31] code. Details on the method are found in the Supplemental Material [24]. Calculation of the frequency-dependent dielectric constant for the C/Ni system relative to the pure Ni metal indicate minimal optical absorption in the C relative to that in the Ni metal (see Supplemental Material [24]) so that all electronic structure changes of the C is a result of excitations in the Ni substrate. For the optical fluence used, the 2T model gives a peak electronic temperature of $T_e \approx 5000$ K within ~100 fs of the optical laser excitation, followed by a surface equilibration to $T \approx 1000$ K after ~2 ps via electron-phonon coupling (for details see Supplemental Material [24]). By comparing with measurements on other transition metals [32,33], we assume that thermalization of the hot electrons and holes occurs on a time scale of ~100 fs.

Fig. 4(a) shows the calculated XAS from the ground state and excited systems corresponding to a high electronic temperature $T_e$ = 5000 K and a fully thermalized system at $T$ = 800 K. The effect of high $T_e$ on the calculated XAS is twofold. The absorption onset, which at 0 K can be identified with $E_F$, is shifted (see S3). In addition, the broad Fermi-Dirac distribution at high $T_e$ means that states well above $E_F$ are populated, thus lowering the XAS intensity which probes unoccupied states. Since we have populated electronic states above the Fermi level, as indicated by the XAS, we should also observe these in the occupied states as probed by XES. Indeed the 5000 K electron distribution of the XES spectra shown in fig. 4(b) shows intensity appearing above the Fermi level.

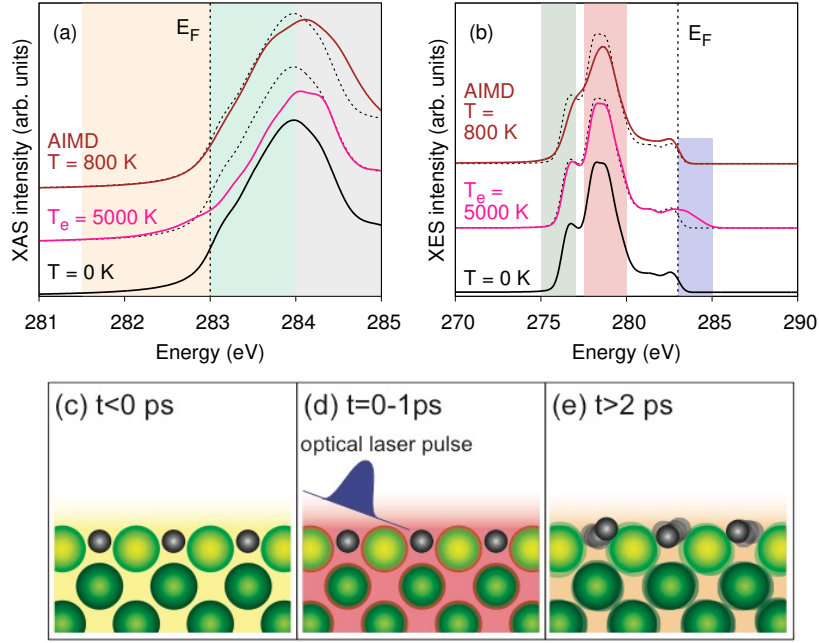

FIG. 4. Computed XAS (a) and XES (b) spectra. The black lines show the vibrational ground state spectra; the brown line the result from an *ab initio* molecular dynamics (AIMD) simulation at 800 K; and the pink line the result of a high electronic temperature. (c-e) Schematic summary of the findings. (c) At t<0, before the optical laser excitation, carbon atoms (black spheres) are adsorbed in hollow sites on the nickel (100) substrate, in which a lattice of positively charged nickel ions (green spheres) share free electrons (shaded yellow region). (d) At t=0, the optical laser pulse excites electrons in the metal to produce hot electrons (shaded red region). (e) The hot electrons lose their energy to substrate phonons and raise the temperature of the adsorbate-substrate system.

The appearance of intensity below $E_F$ and a decreased intensity above $E_F$ is consistent with the experimental XAS at early time as shown in Fig. 2(a). We therefore find that the features of the experimental XAS at t = 0.2 ps are all qualitatively reproduced in the theoretical spectra by simply considering the hot electron-hole (e-h) pair distribution on the C. While the observation (and thermalization) of a hot e-h pair-distribution in bulk metals has been extensively studied via two-photon photoemission studies [32,33] and time-resolved XAS [14,34], here we show the direct observation of this in an adsorbate.

The increase at the high-energy side of the main XAS peak at 4 ps delay is qualitatively reproduced in the simulation of the spectrum from a fully thermalized system at 800 K, as shown by the curve labeled AIMD in Fig. 4(a). The increasing intensity in this region is consistent with thermal motion of C atoms towards lower coordinated bonding geometries than the four-fold hollow sites they occupy initially (see S4). We observe that the excited e-h population seen in the experimental XAS decays on a time scale of ~ 300 fs, simultaneous with an increase in the intensity ~ 1 eV above $E_F$. Thus, these spectral changes are consistent with a decreasing electronic temperature via electron-phonon coupling and corresponding heating of the phonon modes, and is in good agreement with the time scales of the 2T model for Ni (Supplementary Figure S2). In the XES we also observed a similar decay of the intensity around $E_F$ but it does not return to the intensity prior to the pump. This is most likely related to the broadening and shift of the Fermi level at the equilibrated 800 K temperature.

In addition to these effects, the experimental XES spectrum shows some redistribution of intensities in the main peak (Fig. 2(b-c)) at 0.3 ps delay, that cannot be explained by simply a high $T_e$ since these changes occur far below $E_F$. They decay on similar time scales as the holes and electrons in the XAS, indicating that these XES changes are also related to the high electronic temperature. One possibility is that extremely strong non-adiabatic coupling occurs to the parallel vibration of C on the Ni lattice (see Supplemental Material [24]). Ultrafast vibrational excitation (~ 0.1 ps) has been observed previously [11,35,36], although this is not well explained through conventional coupling via electronic frictions which predicts slower vibrational excitations. Another possibility is that many-body or demagnetization effects [37–39] in this highly excited electronic system could affect the width and position of the resonance, and ultimately cause the main XES line shape changes far from $E_F$ in the short time regime. The XES depends sensitively on the 2-dimensional band structure [40], where intensity of C character could be redistributed between different critical points in the Brillouin Zone to cause the XES spectral change, and as the simulations indicate parallel C excitation does induce such a redistribution.

In conclusion, we have obtained XAS and XES of an atomic surface adsorbate with a well understood ground state electronic structure using a pump-probe scheme with an intense femtosecond 400 nm laser pulse as pump and a femtosecond X-ray pulse as probe. The simultaneous measurement of X-ray absorption and emission spectra reveals details of the dynamics on distinct timescales, as summarized in Fig. 4 (c-e). We see a direct effect on both XAS and XES of the initially high electronic temperature, which manifests itself as clearly identifiable changes in the line shape close to the Fermi level and only persists during the first ps. We can identify this as the timescale of thermal electronic excitation. Our observation constitutes direct experimental evidence that ultrafast laser excitation immediately leads to a highly excited e-h pair distribution not just in the substrate, but also in the adsorbate. The XAS high-energy shoulder at longer delays is a clear signature of a high overall temperature, and its gradual build-up indicates equilibration of the system, which takes place over several ps. We identify this timescale as that of phonon excitation and the response of all substrate degrees of freedom to the excitation pulse. We also observe an immediate short time redshift of the main XES peak where the mechanism is more elusive. It could be due to ultrafast excitation of in-plane adsorbate vibrations but also many-body effects in the electronic system itself. This would give a third timescale associated with non-thermal electron-hole pairs and highly selective adsorbate excitation far from equilibrium conditions. Our results underscore the importance of taking high electronic temperature into account when studying e.g. adsorbate dynamics or excitation spectra within a few ps after intense optical pumping. Furthermore, the impact of how the adsorbate electron systems respond to optical stimuli could bring insights into photothermal catalysis, as recently observed for Ni nanoparticles on $SiO_2$ for sustainable fuel production [41].

## ACKNOWLEDGEMENTS

This research was supported by the U.S. Department of Energy, Office of Science, Office of Basic Energy Sciences, Chemical Sciences, Geosciences, and Biosciences Division, Catalysis Science Program to the Ultrafast Catalysis FWP 100435 at SLAC National Accelerator Laboratory under Contract No. DE-AC02- 76SF00515, Knut and Alice Wallenberg Foundation under Grant No 2016.0042, the Swedish Research Council under Grant No 2013-8823, U.S. This research used resources of the National Energy Research Scientific Computing Center, a DOE Office of Science User Facility supported by the Office of Science of the U.S. Department of Energy under Contract No. DE-AC02-05CH11231. The authors acknowledge the continuous support of the FERMI team during the setting up and operation of the FEL source for the experiment. M.B and P.S.M acknowledge funding from the Helmholtz association (VH-NG-1005). M.D.A and R.C. acknowledge support from the SIR grant SUNDYN [Nr RBSI14G7TL, CUP B82I15000910001] of the Italian MIUR. Part of calculations were performed using resources provided by the Swedish National Infrastructure for Computing (SNIC) at the HPC2N and NSC centers. We acknowledge valuable discussions with Lars G. M. Pettersson.

S. S. and E. D. contributed equally to this work.

# REFERENCES

[1] F. Budde, T. F. Heinz, M. M. T. Loy, J. A. Misewich, F. De Rougemont, and H. Zacharias, Phys. Rev. Lett. **66**, 3024 (1991).

[2] H. Petek, H. Nagano, M. J. Weida, and S. Ogawa, J. Phys. Chem. B **105**, 6767 (2001).

[3] M. Bonn, C. Hess, S. Funk, J. H. Miners, B. N. J. Persson, M. Wolf, and G. Ertl, Phys. Rev. Lett. **84**, 4653 (2000).

[4] S. I. Anisimov, B. L. Kapeliovich, and T. L. Perel'man, Zh. Eksp. Teor. Fiz. **66,** 776 (1974) [Sov. Phys.-JETP **39**, 375 (1974)].

[5] C. Frischkorn and M. Wolf, Chem. Rev. **106**, 4207 (2006).

[6] M. Bonn, S. Funk, C. Hess, D. N. Denzler, C. Stampfl, M. Scheffler, M. Wolf, and G. Ertl, Science **285**, 1042 (1999).

[7] M. Dell'Angela et al., Science **339**, 1302 (2013).

[8] H.-Y. Wang et al., Phys. Chem. Chem. Phys. **22**, 2677 (2020).

[9] J. Larue et al., J. Phys. Chem. Lett. **8**, 3820 (2017).

[10] H. Öström et al., Science **347**, 978 (2015).

[11] E. Diesen et al., Phys. Rev. Lett. **127**, 016802 (2021).

[12] M. Wagstaffe et al., ACS Catal. **10**, 13650 (2020).

[13] C. Stamm et al., Nat. Mater. **6**, 740 (2007).

[14] D. J. Higley et al., Nat. Commun. **10**, 5289 (2019).

[15] B. Hammer and J. K. Nørskov, Nature **376**, 238 (1995).

[16] A. Nilsson, L. G. M. Pettersson, B. Hammer, T. Bligaard, C. H. Christensen, and J. K. Nørskov, Catal. Letters **100**, 111 (2005).

[17] A. Nilsson and L. G. M. Pettersson, Surf. Sci. Rep. **55**, 49 (2004).

[18] E. O. F. Zdansky, A. Nilsson, H. Tillborg, O. Björneholm, N. Mårtensson, J. N. Andersen, and R. Nyholm, Phys. Rev. B **48**, 2632 (1993).

[19] P. Finetti et al., Phys. Rev. X **7**, 021043 (2017).

[20] E. Allaria et al., Nat. Photonics **6**, 699 (2012).

[21] J. H. Onuferko, D. P. Woodruff, and B. W. Holland, Surf. Sci. **87**, 357 (1979).

[22] M. Bader, C. Ocal, B. Hillert, J. Haase, and A. M. Bradshaw, Phys. Rev. B **35**, 5900 (1987).

[23] A. Nilsson and N. Mårtensson, Phys. Rev. Lett. **63**, 1483 (1989).

[24] See Supplemental Material at [URL] for details on optical absorption in C/Ni; the two-temperature model; experimental methods; computational methods; XAS from different adsorbate geometries; and discussion of selective vibrational excitation, which

includes Refs. [42-74].

[25] A. Nilsson, J. El. Spect. Rel. Phen. **126**, 3 (2002).

[26] A. Nilsson, O. Björneholm, E. O. F. Zdansky, H. Tillborg, N. Mårtensson, J. N. Andersen, and R. Nyholm, Chem. Phys. Lett. **197**, 12 (1992).

[27] P. Giannozzi et al., J. Phys. Condens. Matter **21**, 395502 (2009).

[28] P. Giannozzi et al., J. Phys. Condens. Matter **29**, 465901 (2017).

[29] M. Taillefumier, D. Cabaret, A.-M. Flank, and F. Mauri, Phys. Rev. B **66**, 195107 (2002).

[30] C. Gougoussis, M. Calandra, A. P. Seitsonen, and F. Mauri, Phys. Rev. B **80**, 075102 (2009).

[31] O. Buñau and M. Calandra, Phys. Rev. B **87**, 205105 (2013).

[32] E. Knoesel, A. Hotzel, and M. Wolf, Phys. Rev. B **57**, 12812 (1998).

[33] M. Lisowski, P. A. Loukakos, U. Bovensiepen, J. Stähler, C. Gahl, and M. Wolf, Appl. Phys. A **78**, 165 (2004).

[34] E. Principi et al., Struct. Dyn. **3**, 023604 (2016).

[35] K.-i. Inoue, K. Watanabe, and Y. Matsumoto, J. Chem. Phys. **137**, 024704 (2012).

[36] K.-i. Inoue, K. Watanabe, T. Sugimoto, Y. Matsumoto, and T. Yasuike, Phys. Rev. Lett. **117**, 186101 (2016).

[37] E. Beaurepaire, J.-C. Merle, A. Daunois, and J.-Y. Bigot, Phys. Rev. Lett. **76**, 4250 (1996).

[38] B. Koopmans, G. Malinowski, F. Dalla Longa, D. Steiauf, M. Fähnle, T. Roth, M. Cinchetti, and M. Aeschlimann, Nat. Mater. **9**, 259 (2010).

[39] B. Y. Mueller, T. Roth, M. Cinchetti, M. Aeschlimann, and B. Rethfeld, New J. Phys. **13**, (2011).

[40] T. Wiell, J. Klepeis, P. Bennich, O. Björneholm, N. Wassdahl, and A. Nilsson, Phys. Rev. B **58**, 1655 (1998).

[41] M. Cai et al., Nat. Energy **6**, 807 (2021).

[42] M. Gajdoš, K. Hummer, G. Kresse, J. Furthmüller, and F. Bechstedt, Phys. Rev. B - Condens. Matter Mater. Phys. **73**, 045112 (2006).

[43] G. K. White, Int. J. Thermophys. **9**, 839 (1988).

[44] Z. Lin, L. V. Zhigilei, and V. Celli, Phys. Rev. B - Condens. Matter Mater. Phys. **77**, 075133 (2008).

[45] D. Novko, J. C. Tremblay, M. Alducin, and J. I. Juaristi, Phys. Rev. Lett. **122**, 016806 (2019).

[46] A. P. Caffrey, P. E. Hopkins, J. M. Klopf, and P. M. Norris, Nanoscale Microscale Thermophys. Eng. **9**, 365 (2005).

[47] S.-S. Wellershoff, J. Güdde, J. Hohlfeld, J. G. Müller, and E. Matthias, Proc. SPIE **3343**, 378 (1998).

[48] I. A. Abrikosov, A. V. Ponomareva, P. Steneteg, S. A. Barannikova, and B. Alling, Curr. Opin. Solid State Mater. Sci. **20**, 85 (2016).

[49] W. You et al., Phys. Rev. Lett. **121**, 077204 (2018).

[50] P. Scheid, G. Malinowski, S. Mangin, and S. Lebègue, Phys. Rev. B **99**, 174415 (2019).

[51] T. Katayama et al., J. El. Spect. Rel. Phen. **187**, 9 (2013).

[52] L. Triguero, L. G. M. Pettersson, and H. Ågren, Phys. Rev. B **58**, 8097 (1998).

[53] E. Diesen, G. L. S. Rodrigues, A. C. Luntz, F. Abild-Pedersen, L. G. M. Pettersson, and J. Voss, AIP Adv. **10**, 115014 (2020).

[54] A. Nilsson et al., Chem. Phys. Lett. **675**, 145 (2017).

[55] H. Öberg et al., Surf. Sci. **640**, 80 (2015).

[56] B. Hammer, L. B. Hansen, and J. K. Nørskov, Phys. Rev. B **59**, 7413 (1999).

[57] S. Mallikarjun Sharada, R. K. B. Karlsson, Y. Maimaiti, J. Voss, and T. Bligaard, Phys. Rev. B **100**, 035439 (2019).

[58] D. Vanderbilt, Phys. Rev. B **41**, 7892 (1990).

[59] M. Leetmaa, M. P. Ljungberg, A. Lyubartsev, A. Nilsson, and L. G. M. Pettersson, J. El. Spect. Rel. Phen. **177**, 135 (2010).

[60] A. Föhlisch, J. Hasselström, P. Bennich, N. Wassdahl, O. Karis, A. Nilsson, L. Triguero, M. Nyberg, and L. Pettersson, Phys. Rev. B - Condens. Matter Mater. Phys. **61**, 16229 (2000).

[61] G. Kresse and J. Furthmüller, Phys. Rev. B **54**, 11169 (1996).

[62] G. Kresse and J. Furthmüller, Comput. Mater. Sci. **6**, 15 (1996).

[63] G. Kresse and D. Joubert, Phys. Rev. B - Condens. Matter Mater. Phys. **59**, 1758 (1999).

[64] P. E. Blöchl, O. Jepsen, and O. K. Andersen, Phys. Rev. B **49**, 16223 (1994).

[65] J. J. Mortensen, L. B. Hansen, and K. W. Jacobsen, Phys. Rev. B **71**, 035109 (2005).

[66] J. Enkovaara et al., J. Phys. Condens. Matter **22**, 253202 (2010).

[67] S. Flügge, *Practical Quantum Mechanics* (Springer-Verlag Berlin Heidelberg, 1971).

[68] S. Baroni, S. de Gironcoli, A. Dal Corso, and P. Giannozzi, Rev. Mod. Phys. **73**, 515 (2001).

[69] M. Calandra and F. Mauri, Phys. Rev. B **71**, 064501 (2005).

[70] M. Wierzbowska, S. de Gironcoli, and P. Giannozzi, ArXiv Prepr. Cond-Mat/0504077 (2005).

[71] D. Menzel and R. Gomer, J. Chem. Phys. **41**, 3311 (1964).

[72] P. A. Redhead, Can. J. Phys. **42**, 886 (1964).

[73] J. A. Misewich, T. F. Heinz, and D. M. Newns, Phys. Rev. Lett. **68**, 3737 (1992).

[74] J. Gavnholt, T. Olsen, M. Engelund, and J. Schiøtz, Phys. Rev. B **78**, 075441 (2008).

# SUPPLEMENTAL MATERIAL

# CONTENTS

# I. OPTICAL ABSORPTION IN C/Ni

The frequency-dependent dielectric matrix of p4g-C/Ni(100) and clean Ni(100) (with p4g surface layer rotations) were calculated in an independent particle picture from the DFT ground state electronic structures [1]. Using these dielectric tensors, the optical absorption (1-R) for each surface was calculated using Fresnel's equations for the conditions of the experiments (3 degrees angle of incidence with P polarized light). As can be seen in the Figure S1, the optical absorption in C/Ni is nearly indistinguishable from that of Ni and emphasizes that nearly all of the optical absorption for C/Ni occurs in the Ni rather than direct optical excitation of the Ni 3d – C 2p bands.

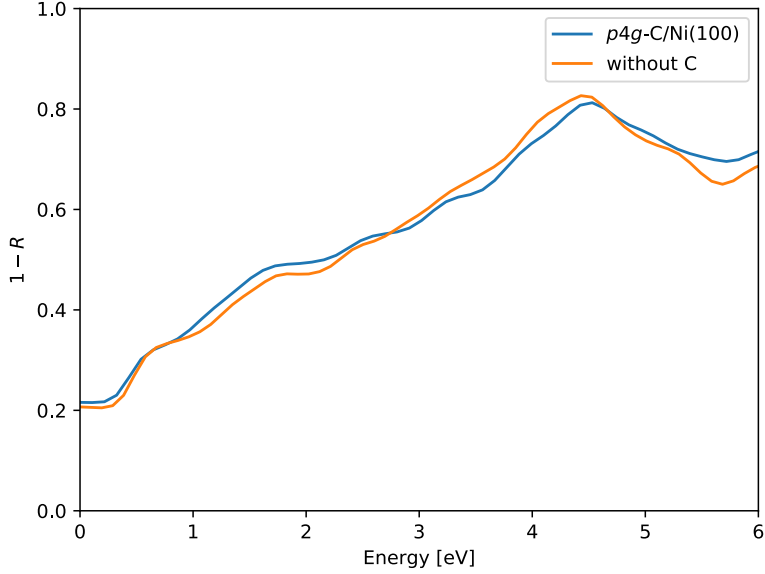

**Fig. S1:** Optical absorption in C/Ni(100) compared to a clean Ni(100) surface.

# II. TWO-TEMPERATURE MODEL

The common two-temperature (2T) model [2] describes the electrons and phonons as two coupled heat baths with depth- and time-dependent temperatures $T_e, T_{ph}$ whose time evolution is governed by the equations:

$$C_e \frac{\partial T_e}{\partial t} = \frac{\partial}{\partial z}\left(\kappa \frac{\partial T_e}{\partial z}\right) - g(T_e - T_{ph}) + S(z,t)$$

$$C_{ph} \frac{\partial T_{ph}}{\partial t} = g(T_e - T_{ph})$$

$$S(z,t) = I(t)e^{-z/\lambda}/\lambda$$

The heat capacity of the phonons is calculated using the Debye model:

$C_{ph} = 9nk_B \left(\frac{T_{ph}}{\Theta_D}\right)^3 \int_0^{\Theta_D/T_{ph}} \frac{x^4 e^x}{(e^x-1)^2}$, where $\Theta_D$ is the Debye temperature of the metal and $n$ the atomic density. The thermal conductivity is written as $\kappa = \kappa_0 \frac{T_e}{T_{ph}}$, with $\kappa_0$ taken as the low temperature value.

The laser pulse is modeled with a Gaussian time profile $I(t) = F\exp(-\frac{t}{2\sigma^2})/\sqrt{2\pi\sigma^2}$, where $F$ is the absorbed fluence on the sample and $\sigma = \Gamma/2.355$ with $\Gamma$ the full width at half maximum (FWHM). The initial temperature was set to 70 K. All other constant parameters are tabulated below. The 2T model is for a pure Ni surface and does not include the modest changes due to the adsorbed C.

The 2T model usually considers a temperature-independent electron-phonon coupling factor $g$, and a simple linear T-dependence of the electron heat capacity $C_e = \gamma T_e$. For Ni a marked departure from this simple behavior has been both observed [3] and found in theoretical studies [4]. We therefore allow a temperature dependent $g(T)$ and $C_e(T)$, using DFT calculations to obtain numerical values for these two quantities (see below).

The resulting temperatures in the surface layer are shown in Fig. S2. We have used the traditional 2T model which only considers the phonon modes of the metal. Adding the phonon modes of the adsorbate C can shorten the time for the $T_{ph}$ rise [5]. However, their inclusion does not affect the rise or peak in $T_e$.

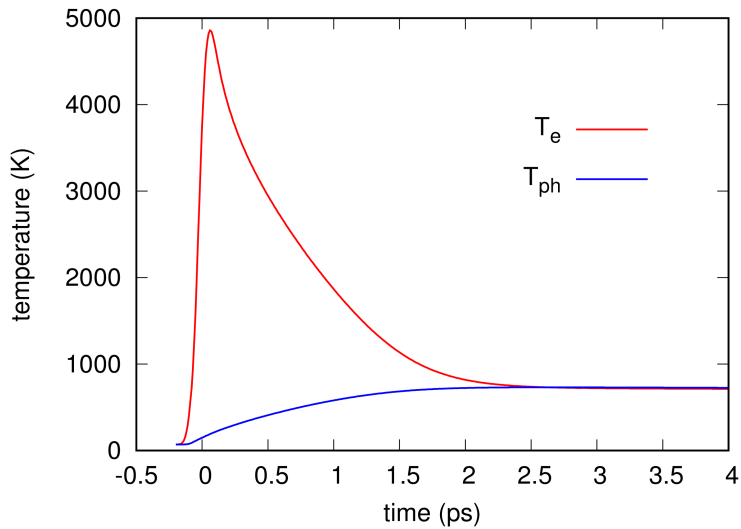

**Fig. S2:** Results from the 2T model with T-dependent parameters.

| $n$ | $9.13 \cdot 10^{28}$ m$^{-3}$ | Atomic density |
|---|---|---|
| $\Theta_D$ | 477.4 K | Debye temperature of Ni |
| $\kappa_0$ | 94 Wm$^{-1}$K$^{-1}$ | Electron heat conductivity |
| $F$ | 126 Jm$^{-2}$ | Absorbed fluence |
| $\lambda$ | 13.49 nm | Optical penetration depth |
| $\Gamma$ | 100 fs | Pulse duration |

**Table SI:** Parameters used in the 2T model

**Properties of Ni**

To accurately model the thermal response of the bulk Ni we use a temperature-dependent electronic specific heat $C_e(T)$ and electron-phonon coupling $g(T)$. In particular for $g$, there are significantly different values in the literature [6–8] due to the separate temperature ranges where different measurement techniques can be applied. To obtain $g(T)$ and $C_e(T)$ as functions of temperature, we calculate the bulk Ni DOS and then follow the approach of Ref. 4. For a given DOS $\rho(E)$, the temperature-dependent Fermi level $\mu(T)$ is given by solving the implicit equation

$$N_e = \int f(E,\mu(T),T)\rho(E)dE$$

where $f$ is the Fermi-Dirac distribution and $N_e$ the number of electrons. We then calculate the heat capacity and electron-phonon coupling as

$$C_e(T) = \frac{\partial U_e}{\partial T} = \frac{\partial}{\partial T}\int f(E,\mu(T),T)\rho(E)E dE$$

$$g(T) = \frac{\pi\hbar k_B \lambda\langle\omega^2\rangle}{\rho(E_F)}\int \rho^2(E)\left(-\frac{\partial f}{\partial E}\right)dE \qquad (S1)$$

where $U_e$ is the internal energy of the electronic system. The quantity $\lambda\langle\omega^2\rangle$ is obtained by matching $g(T)$ at room temperature to experimental data [7].

These quantities were calculated in Ref. 4 using spin-paired calculations for Ni. While this is a crude, but reasonable approximation at temperatures far above the Curie temperature $T \gg T_c$, where the DOS of the paramagnetic state should resemble that of spin-paired Ni and the obtained $C_e(T)$ agrees quite well with experiment (Fig. S3c, experimental results from Ref. 3), for room temperature quantities the DOS from spin-polarized calculations is more accurate so as to avoid the unphysical sharp peak at $E_F$ in the spin-paired DOS (cf. Fig. S3a). Around $T_c$ there should be a smooth transition between the spin-polarized and spin-paired situations. Spin-polarized DFT without treatment of the randomly oriented paramagnetic moments cannot properly account for the paramagnetic phase nor the magnetic phase transition at the experimental Curie temperature $T_c = 631$ K, but rather finds a ferromagnetic ground state up to $T \approx 3000$ K. A complete theoretical treatment of this temperature region requires a treatment of the disordered magnetic moments which is beyond the scope of this work [9]. Note that ultrafast demagnetization [6,10] happens within 20 fs of the driving pulse [11]. Here we take the pragmatic approach of simply dividing the temperature range at the experimental $T_c$. Below this, we use the spin-polarized DOS (summed over the two spins) to calculate $g(T)$ and $C_e(T)$; above it we use the spin-paired DOS. This gives unphysical discontinuities at $T_c$. Since we are however only using our results for the relatively crude 2T model, we find this inaccuracy to be less severe than resorting to a completely spin-paired treatment as in Ref. 4, or using DFT for the full temperature range as in [12], where ferromagnetism persists up to 3000 K giving significant deviations in $C_e(T)$ when compared to experiment. For $C_e(T)$ inclusion of spin has little effect on the results; since we reach temperatures $T \approx 5000$ K, some deviations from the correct value for $T < T_c$ will be of little consequence. For $g(T)$ it is more sensitive since the experimental room temperature value of $g$ is used to extract the unknown quantity $\lambda\langle\omega^2\rangle$ in Eq. S1. For this purpose we use the spin-polarized DOS, giving us a value of $\lambda\langle\omega^2\rangle = 91.7$ meV$^2$, compared to $\lambda\langle\omega^2\rangle = 49.5$ meV$^2$ in Ref. 4 where the spin-paired DOS was used also at room temperature. This leads to a stronger electron-phonon coupling, which still compares well with available experiments [6,7]. (Note that, in Ref. 8, a linear $C_e(T)$ was assumed, leading to a significantly underestimated $T_e$ and similarly underestimated $g$.) It is however clear that these parameters are obtained through crude approximations, and the 2T results should therefore only be taken as a rough estimate of the actual temperature evolution of the substrate.

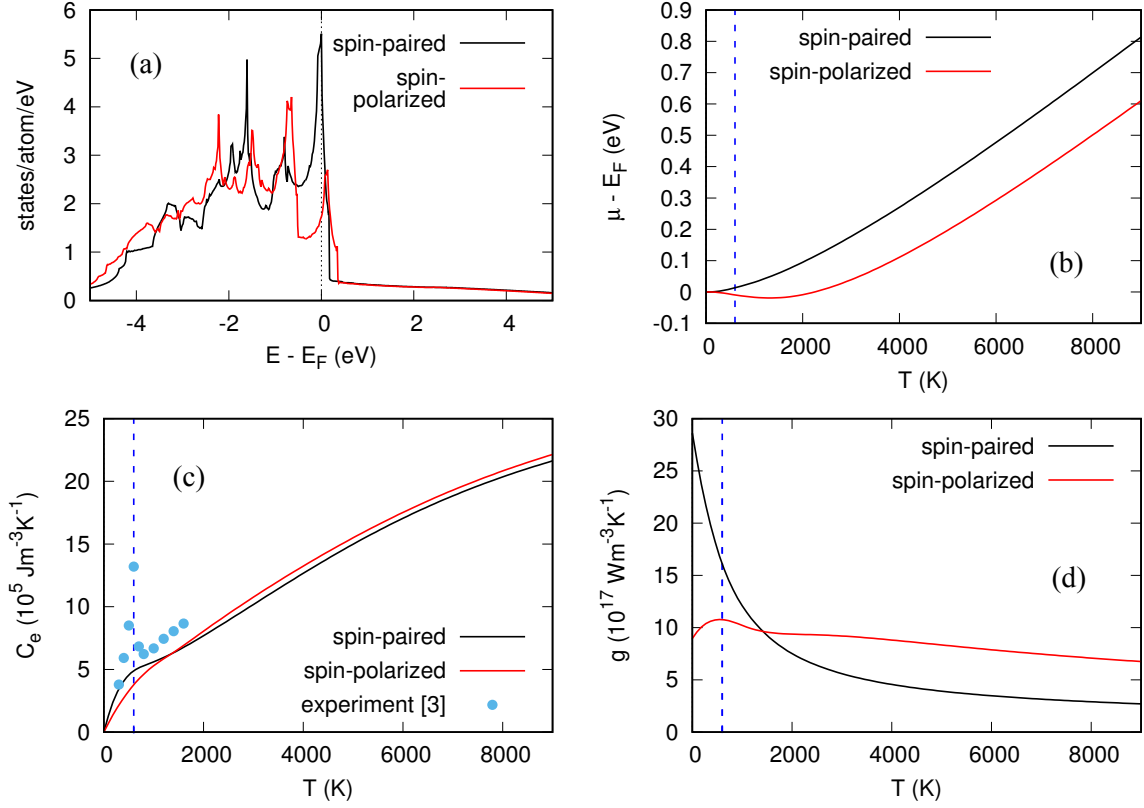

**Fig. S3:** Modelling of high temperature Ni behavior, comparing spin-polarized to unpolarized results. (a) Densities of states (summed over the two spins); (b-d) temperature dependence of the Fermi level, the electronic specific heat, and the electron-phonon coupling g, respectively. The dashed lines indicate the experimental Curie temperature $T_c$; the experimental $C_e(T)$ values are taken from Ref. 3.

## III. EXPERIMENTAL METHODS

**Experimental Conditions**

The optical laser pump and soft X-ray laser probe measurements were performed at the DiProI beamline of the FERMI free-electron laser (FEL) in Trieste, Italy [13]. We used 400 nm optical laser pulses with a pulse duration of 90-100 fs that were generated from the same laser system that is used to seed the FERMI FEL, and soft X-ray pulses in the carbon K-edge photon energy range around 290 eV (4 nm wavelength) with a pulse duration of 45-50 fs. Since seed and pump pulses were generated from the same laser system, the arrival times of the optical laser pump and the FEL X-ray pulse at the sample are intrinsically synchronized. The combined time profile results in an overall resolution of 100-110 fs determined solely by the pulse durations of optical laser and FEL. For X-ray absorption measurements we set the central average FEL photon energy to a number of pre-defined values covering the desired range of the carbon K-edge X-ray absorption resonance with a step size of 0.3 – 0.5 eV. For each photon energy the delay between optical laser and FEL pulse was set to pre-defined values recording 1000 FEL shots for each delay. For X-ray emission measurements we set the central average FEL photon energy to the desired value and varied the delay between optical laser and FEL pulse between pre-defined values recording 1000 FEL shots for each setting. Each delay setting was revisited 10-20 times.

The pump laser had a spot size of ~150×150 µm$^2$ and a pulse energy of ~152 µJ, and the X-ray probe beam had a spot size of ~30×30 µm$^2$. Both the laser and the x-ray beam had a grazing incidence angle, which was estimated to be 2.24°.

The XAS spectra were recorded by stepwise varying the incidence energy and detecting fluorescence using a large area detector with a perylene filter to remove background UV light. The XES spectra were recorded with a soft X-ray spectrometer similar to the spectrometer previously used at LCLS [14] with an overall resolution of 0.3 eV.

**Sample Preparation**

We used established surface science techniques to cover a clean Ni(100) surface with 0.5 ML of carbon that form a p4g(2x2) surface structure. The sample was a standard commercially available Ni(100) single-crystal (from Surface Preparation) with a round surface of 20 mm diameter. The sample was cleaned and prepared in ultra-high vacuum (UHV) condition at a base pressure of <5×10$^{-10}$ mbar. Prior to the experiments initial cleaning of the sample was performed by several tens of cycles of 10 minutes Neon$^+$ sputtering at room temperature followed by 30 minutes of annealing at 600 K. During the experiment the cleaning cycles consisted of Neon$^+$ sputtering at 470 K for 10 minutes followed by flash annealing the sample to 1100 K. After the cleaning procedures, the carbon coverage was prepared by dosing 40 L of C$_2$H$_2$ at 470 K and thereafter flashing the sample to 650 K to dissociate the C$_2$H$_2$ into free carbon atoms on the surface. With this procedure a half monolayer coverage of C on the surface was achieved. After sample preparation the sample was cooled down to liquid nitrogen temperature by a constant liquid nitrogen flow through the cryostat and the experimental chamber was pumped down to its base pressure <5e-10 mbar. No background pressure was applied. Even though the optical laser excitation does not result in irreversible changes of the surface structure we continuously scanned the sample through the laser and FEL beam providing a fresh spot on the sample for each laser and FEL shot-pair. This was done as a precaution to avoid long term heating and possible melting of the substrate. Each spot on the sample was revisited after a cycle period of 10-20 seconds.

**Analysis**

In the post-processing of all recorded data we analyzed the FERMI online spectrometer on a shot-to-shot basis. This spectrometer provides the spectral intensity distribution for each FEL shot, which allowed us to assign an actual central photon energy, bandwidth and intensity for each shot. The analysis showed that, for a fixed average central photon energy, the shot-to-shot fluctuations of the actual central photon energy was ~0.7 eV FWHM and the single-shot FEL bandwidth varied between 0.2 and 1.0 eV FWHM. We used the single-shot values to filter out shots with low intensity or large FEL bandwidth (FWHM>0.6 eV). For X-ray emission measurements we filtered out in addition shots for which the actual central photon energy was off from the desired photon energy by more than ±0.15 eV. For X-ray absorption measurements we resorted all shots based on their actual central photon energy into bins of 0.25 eV width. For each bin the accumulated Total Fluorescence Yield (TFY) intensity was normalized by the accumulated incident intensity and the standard error was calculated based on Poisson statistics.

The transient changes in XAS and XES intensity in Figure 3 were modeled with a sum of two Gaussian error functions of the form $I(t) = a \cdot \int_{-\infty}^{t} e^{-(t-t_0)/\sigma^2}$. We used Gaussian fitting for the rise and decay in order to mimic the resolution of the convoluted two laser pulses and to observe if there are any additional delays beyond the instrumental time broadening. Only the blue delay trace in Figure 3(a) was modeled by a single Gaussian error function. We denote the value for $\sigma$ as the characteristic rise or decay time and $t_0$ as the edge position representing the inflection point of the rising or falling flank.

# IV. COMPUTATIONAL METHODS

**Spectrum calculations at equilibrium positions**

To identify the dynamics causing changes in the XAS and XES, we calculate X-ray spectra using GGA-level density functional theory (DFT) calculations and the transition-potential (TP) approach [15]. This method has been successfully used to identify detailed structural changes in adsorption via X-ray spectroscopy [16] and for interpretation of the ultrafast evolution of X-ray spectra in terms of underlying dynamical features [17–20].

We mainly use the DFT code Quantum ESPRESSO [21,22], which uses a plane-wave basis for the valence electrons and pseudopotentials to represent the core electrons. We use the RPBE [23] functional due to its accuracy in determining surface adsorption energies [24]. All calculations are spin-polarized (unrestricted Kohn-Sham). We use a four-layer 2x2 slab model for initial geometry optimization, with C/Ni adsorbed in four-fold hollow sites at ½ ML coverage, with periodic boundary conditions in all directions and 20 Å of vacuum separating the slab from its periodic image. An RPBE optimimum lattice constant of 3.553 Å for bulk fcc-Ni was used, and a 4x4x1 Monkhorst-Pack grid for Brillouin zone integration. Optimizing the atomic positions correctly predicts the p4g surface restructuring upon adsorption of C on the Ni(100) surface. For the excited-state calculations, we double the cell in the x- and y-directions to 4x4x4 (correspondingly reducing the k-point grid to 2x2x1) to reduce the interaction of the core hole with its periodic images. XAS spectra were computed using a refined 11x11x1 k-point grid, while keeping the density fixed (non-SCF/Harris calculation). Ultrasoft pseudopotentials [25] were used for atoms without core excitations. The plane-wave and density cutoffs were correspondingly chosen for the surface calculations to be 500 and 5000 eV, respectively. The geometries were optimized until all forces were less than 0.01 eV/Å, with the bottom two layers kept fixed.

The XAS is calculated using a Lanczos recursion Green's function technique, as implemented in the xspectra code [26–28]. For the spectrum calculations, a half-core-hole pseudopotential was used on the excited atom (transition-potential approach [15]) which has been shown to give reliable results [15,29]. For the core-excited C, norm-conserving half and full core-hole pseudopotentials were generated using the 'atomic' code included with the Quantum Espresso distribution [21]. Full and half-core hole pseudopotentials are generated by occupying the 1s shell in the single-atom DFT solution by only, respectively, 1 and 1.5 electrons. The obtained spectra were shifted according to our calculated absorption onsets (Δ-Kohn-Sham method). All spectra were finally shifted by a common reference energy, chosen in order to match the experimental spectrum at negative delay with the calculated spectrum of the adsorbate in its optimized position.

XES is calculated in an analogous way using xspectra, however no Δ-Kohn-Sham procedure is needed since the core binding energy does not enter into the final XES emission energy [30]. Furthermore, no explicit core hole pseudopotential is needed since the final state of XES decay contains a ground state core [31]. For this reason the spectrum calculation can be performed using the 2x2x4 slab model. An overall shift is performed also for the XES to align with the experimental unpumped spectra, and an overall broadening is applied to account for lifetime effects and instrumental resolution. C/Ni does not get significantly vibrationally excited during the XES process [32,33], allowing for the simplified treatment used here without the need for the full Kramers-Heisenberg formalism.

For calculating the parameters of the 2T model, calculations for bulk Ni were performed using the VASP code [34,35] with ultrasoft (PAW) pseudopotentials [36]. An elementary fcc unit cell and a 21x21x21 k-point grid was used, with the Blöchl tetrahedron method [37] to obtain the density of states.

**Spectrum calculations at $T_e$ = 5000 K**

To simulate the effect of a high electronic temperature on XAS and XES, we first obtain the electronic density from an ordinary DFT calculation with the electronic states populated according to a 5000 K Fermi-Dirac distribution. The presence of holes below and electrons above the Fermi level ($E_F$) affects XAS and XES in opposite ways. In XAS the excited core electron can now populate states below $E_F$ that would already be filled at low temperature, increasing the XAS intensity there; above $E_F$ the fact that some states are already filled instead lowers the XAS intensity. In XES the reverse is seen: the intensity above $E_F$ is increased while the intensity below decreases. These effects are included in the spectrum calculation by explicitly including the thermal occupation numbers when calculating the X-ray transition rates.

**Spectrum calculations at T = 800 K**

To simulate the effect of an equilibrated hot system with both electrons and phonons thermalized, corresponding to the situation at $t > 2$ ps, we perform an *ab initio* molecular dynamics (AIMD) simulation using VASP. The cell was identical to the one used in Quantum ESPRESSO for optimization. Starting at the equilibrium geometry with randomized initial velocities corresponding to a Maxwell-Boltzmann distribution at the desired temperature of 800 K, we use a Langevin thermostat to simulate a thermal ensemble. After equilibrating the system for 1 ps, the dynamics was run for another 1 ps with a time step of 2 fs. Every 100 fs, the structure was saved and the XAS and XES were calculated for that static structure using the Quantum ESPRESSO package. These spectra were then averaged together for the final result.

# V. XAS FROM DIFFERENT ADSORBATE GEOMETRIES

To investigate the origin of the additional intensity at energies above the main XAS peak in the experimental 4.0 ps delay spectrum, and the corresponding shift in the spectrum from AIMD, we calculate the XAS at intermediate geometries as the C atom moves from the hollow to a bridge site. At high temperatures, both motion of substrate atoms and of the adsorbate itself will contribute to the C moving out of the hollow site. We calculate the XAS for the p4g geometry and three intermediate geometries (ptb-1, ptb-2 and ptb-3) between p4g and the geometry where one C atom is moved to a nearby bridge site, along the minimum energy path obtained with a nudged elastic band (NEB) simulation. The structures are shown in Figure S4. The displacement of the C atom from the p4g geometry for ptb-1, ptb-2 and ptb-3 are 0.32 Å, 0.90 Å and 1.56 Å, respectively. These calculations were performed using the GPAW [38,39] code; the computational parameters were the same as for other calculations (see Supplement IV). Franck-Condon transitions into vibrationally excited states perpendicular to the surface were taken into account in the following way: For each geometry, the carbon atom is vertically displaced from -0.5 Å to +1.5 Å from the original position, and the potential energy and the XAS are calculated for every displacement point (41 in total). The vibrational eigenstates on the ground and excited PES are then calculated numerically, and the eigenfunction overlap gives the Franck-Condon factors. The vibrational excitations turn out to have limited influence on the spectra, consistent with experimental results [32]. Finally, a Gaussian broadening of 1.05 eV and a Lorentzian broadening of 0.3 eV were applied to the obtained spectra.

All computed theoretical XAS curves are plotted together with the experimental XAS at the negative time delay –0.4 ps in Figure S5. The p4g geometry gives a main peak at around 283.6 eV. When the C atoms moves towards the nearby bridge site, the peak position moves towards higher energies, which corresponds to the increase in intensity in the experimental result at around 284.6 eV.

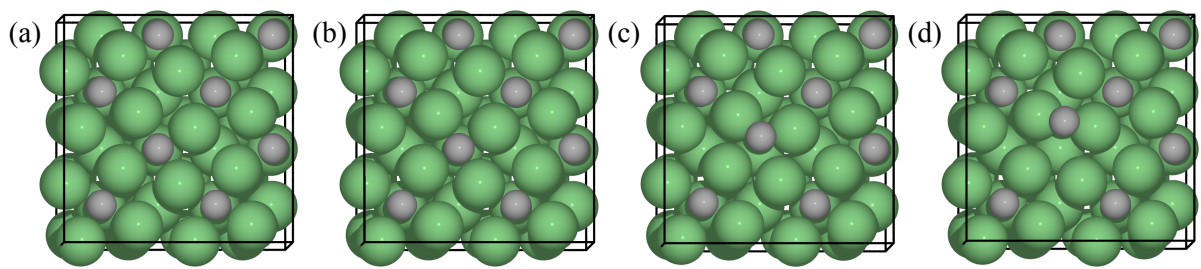

**Figure S4:** Atomic models for (a) p4g C/Ni, the intermediate geometries from p4g to bridge (b) ptb-1, (c) ptb-2, and (d) ptb-3. The green and black spheres represent Ni and C atoms, respectively. The ptb-3 geometry is very close to the bridge geometry but not exactly at the bridge site.

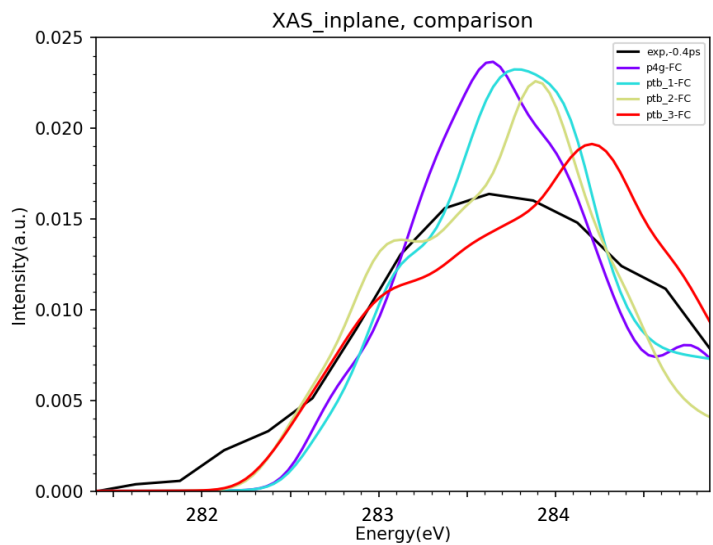

**Figure S5:** Experimental XAS at the delay –0.4 ps, and the computed XAS for the geometries shown in Figure S4.

## VI. SELECTIVE VIBRATIONAL EXCITATION OF ADSORBATES

We propose ultrafast excitation of in-plane C vibrations as a possible explanation for the very rapid redshift observed in the XES. In this section we describe the details of the spectra obtained by selectively exciting particular vibrational modes of the C atoms, and models for energy transfer from the excited electrons of the substrate to these vibrational modes (=phonons). Neither our electronic friction model with calculated electron-phonon couplings nor an ion-resonance model where excited electrons directly populate the antibonding C-Ni resonance can quantitatively reproduce this very fast (< 0.25 ps) and selective excitation of the C in-plane vibration. However, ultrafast vibrational excitation (~100 fs) of adsorbate modes has been observed previously [20,40,41], suggesting that ballistic non-thermalized electrons formed initially by the optical laser excitation of the metal could couple much more strongly to adsorbate motion than the thermalized hot electrons described by an electronic frictional coupling.

**Spectra from vibrationally excited adsorbates**

To calculate the spectrum resulting from a vibrationally excited adsorbate, we follow a similar method as in Ref. 20. We first observe that, in the optimized structure, the C atoms occupy the four-fold hollow sites almost in plane with the top layer Ni atoms. A vibrational analysis allowing only one C atom to vibrate gives two degenerate eigenfrequencies 20.48 THz for in-plane ($x,y$) vibrations, and 9.35 THz for out-of-plane ($z$) vibrations. For obtaining approximate spectra we assume that each C vibrates independently of each other in a cylindrically symmetric potential. Due to the anharmonicity of the potential in z-direction, the potential energy was calculated for several displacements, and a Morse potential $V(z) = D_e(1 - e^{-a(z-z_0)})$ was fitted to the values. The obtained potential is shown in Fig. S6. The vibrational states are then analytically given by:

$$\psi_n^{(gs)}(z) = N_n z^{\lambda - n - \frac{1}{2}} e^{-\frac{z}{2}} L_n^{(2\lambda - 2n - 1)}(z)$$

where $z = 2\lambda e^{-(x-x_0)}$; $x = ar$; $\lambda = \frac{\sqrt{2mD_e}}{a\hbar}$ ($m$ is the mass of C); $N_n = \sqrt{\frac{n!(2\lambda - 2n - 1)}{\Gamma(2\lambda - n)}}$ and $L_n^\alpha(z)$ is a generalized Laguerre polynomial.

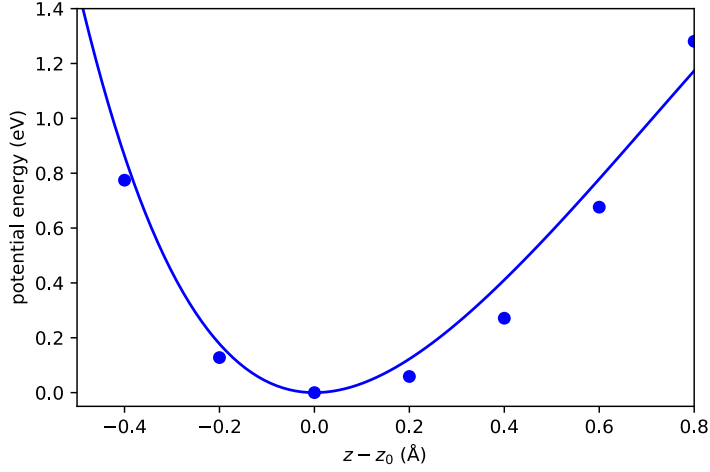

**Fig. S6:** The calculated PES (dots) and Morse potential fit (solid) in the out-of-plane direction

The in-plane vibration was modeled using a radially symmetric quadratic potential corresponding to frequency 20.48 THz in the radial direction, giving a 2D harmonic oscillator. Its eigenstates can be written in terms of the radial coordinates $(r, \varphi)$ as [42]:

$$\chi_{n,M}(r, \varphi) = C_{p,M} \theta^{|M|} e^{-\frac{\lambda r^2}{2}} L_p^{|M|}(\lambda r^2) e^{iM\varphi}$$

where $\lambda = m\omega/\hbar$, $p = (n - |M|)/2$, $C_{p,M} = \sqrt{\frac{2\lambda^{|M|+1} p!}{(p+|M|)!}}$ and $L_p^{|M|}$ is the generalized Laguerre function. $M$ takes the values $-n, -n+2, \ldots, n-2, n$ giving an $n+1$-fold degeneracy of the $n^{th}$ level. The energy levels are given by $E_n = \hbar\omega(n + 1)$, with $n = 0,1,2,\ldots$.

For the perpendicular direction we then calculate the spectra $w(E, z)$ for a number of displacements. We interpolate the result to get a continuous 2D function, and then calculate the spectrum $w_n(E)$ for each vibrational state as:

$$w_n(E) = \int w(E, z) |\psi_n^{(gs)}(z)|^2 dz \qquad (S2)$$

For the in-plane direction we calculate in an analogous way the spectra for radial displacements $w(E, x)$. We approximate the dependence on the azimuthal angle $\varphi$ as $w(E, r, \varphi) \approx w(E, x)\cos^2\varphi + w(E, y)\sin^2\varphi$. For a single state $\chi_{n,M}$ we then get

$$w_{n,M}(E) = \int w(E, r, \varphi) |\chi_{n,M}(r, \varphi)|^2 r d\varphi dr \propto \int (w(E, x) + w(E, y)) |\chi_{n,M}(r)|^2 r dr$$

since $|\chi_{n,M}|^2$ is independent of $\varphi$. Writing $w(E,r) \equiv w(E,x) + w(E,y)$ we then calculate the total spectrum in a similar way to Eq. S2. We need to sum over all allowed $M$-values for each $n$, which yields the following expression

$$w_n(E) = \sum_M \int w(E,r)|\chi_{n,M}(r)|^2 r dr$$

To finally calculate the spectrum at an elevated temperature T, the spectra from the individual states are added together according to the Boltzmann distribution to give:

$w(E,T) = \sum_n P_n(T) w_n(E)$ with $P_n(T) = \dfrac{e^{-E_n/k_B T}}{\sum_n e^{-E_n/k_B T}}$

The resulting spectra are shown in Fig. S7 for T=0 K, 2000 K and 4000 K.

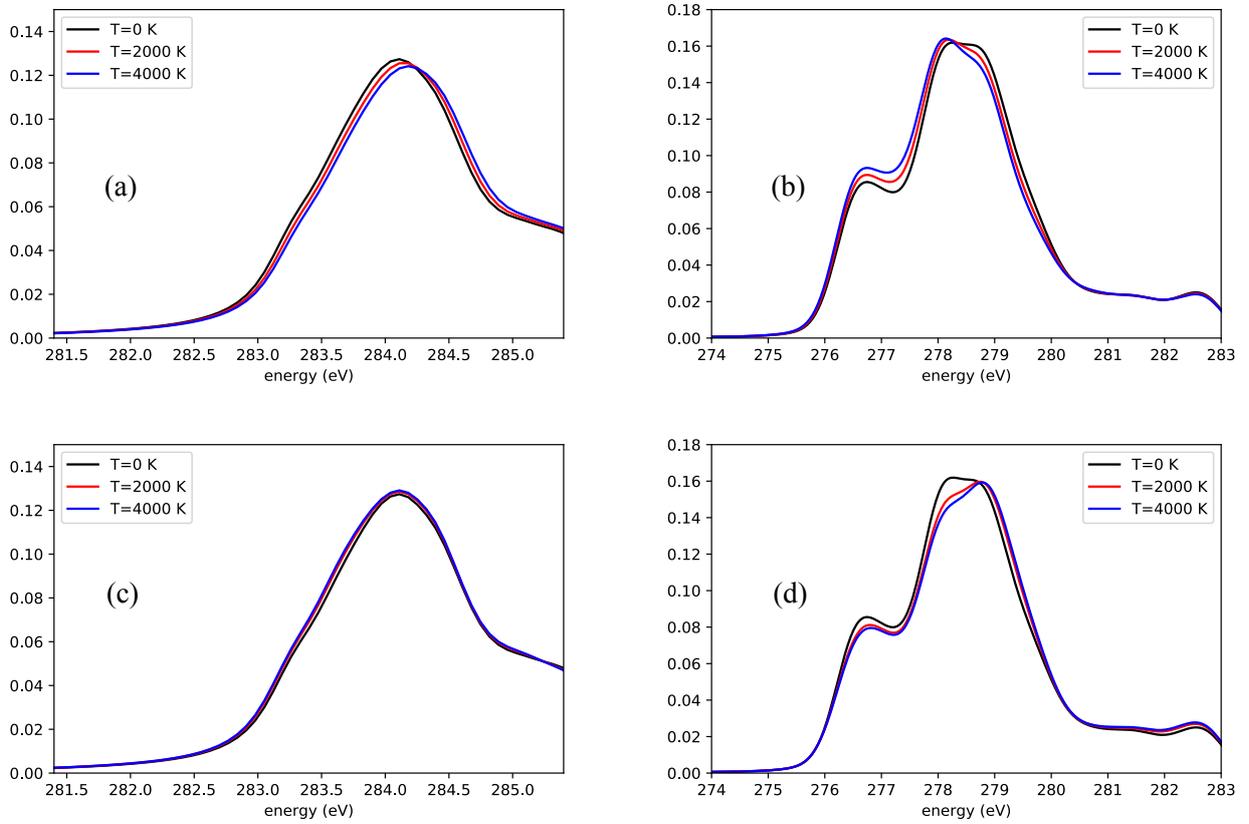

**Fig. S7:** Calculated XAS (left) and XES (right) with vibrationally high temperature. (a) and (b): excited in-plane vibration; (c) and (d); excited out-of-plane vibration.

**Calculation of lifetimes of adsorbate modes by electron-phonon coupling**

We calculate the theoretical lifetimes of the C/Ni(100) adsorbate vibrational modes using the DFPT [43] method implemented in Quantum ESPRESSO. Details on the method and its implementation can be found in e.g. [44,45]. Our approach has been described in detail in Ref. 20. The phonon linewidth $\gamma_{qv}$ is calculated by evaluating the Fermi-golden-rule-like expression

$$\gamma_{qv} = 2\pi\omega_{qv} \sum_{ij} \int \frac{d^3\mathbf{k}}{\Omega_{BZ}} |g_{qv}(\mathbf{k},i,j)|^2 \delta(E_{\mathbf{q},i} - E_F)\delta(E_{\mathbf{k+q},j} - E_F) \qquad (3)$$

where $g_{qv}(\mathbf{k},i,j)$ are the electron-phonon coefficients. The double-delta expression in Eq. 3 is evaluated by replacing the delta functions by Gaussian functions of width σ (for our results, σ=0.6 eV was used). We use a 16x16x1 k-point grid to evaluate the linewidths, using the interpolation scheme described in Ref. 45. Since the electron-hole pair scattering can excite phonons at any $\mathbf{q}$, we calculate $\gamma_{qv}$ for a grid of points in the first Brillouin zone and average the result:

$$\bar{\gamma}_v = \int \frac{d^2\mathbf{q}}{\Omega_{BZ}} \gamma_{qv}$$

from which we get the lifetime as $\tau_v = \hbar/\bar{\gamma}_v$.

The obtained lifetimes of the highest frequency modes are shown in Table SII. The first four modes correspond to in-plane vibrations of the two carbon atoms, while the following two modes are symmetric and asymmetric out-of-plane vibrations. Due to the relatively small mass difference between C and Ni, especially the out-of-plane vibrations are somewhat mixed with substrate vibrations. However the obtained values give at least a semi-quantitative estimate of the relative coupling strengths. The modes with lower frequencies, dominated by substrate vibrations, have lifetimes in the range 4.5-6 ps.

| Lifetime (ps) | Mode character |
|---|---|
| 1.74 | In-plane |
| 1.71 | In-plane |
| 2.65 | In-plane |
| 3.93 | In-plane |
| 3.05 | Out-of-plane |
| 2.39 | Out-of-plane |

**Table SII:** Calculated lifetimes of the six highest frequency modes.

**Ion resonance model for adsorbate excitation**

A common picture for the coupling between excited electrons and adsorbate motion is the ion resonance model, where a single energetic electron resonantly populates an unoccupied orbital of the adsorbate. This is the mechanism behind the Desorption Induced by (Multiple) Electronic Transitions (DIET [46,47] and DIMET [48]) processes. Within this picture, we can get an estimate of the energy transfer to the adsorbate IS mode due to resonant population of the adsorbate LUMO by scattering of excited electrons from the substrate into the resonance. To this end, we calculate the potential energy surface (PES) for the C using the linear-expansion ΔSCF method [49], which ensures that an empty C p-derived orbital is populated by one electron. The GPAW code [38,39] was used for the excited-state PES calculation, using identical parameters to the Quantum ESPRESSO calculations in the rest of this work. To define the orbital to be filled, an SCF calculation is first done removing all atoms except one C from the unit cell. This will give 3 p-bands (one with $p_z$ character and two with $p_{xy}$ character) with weak dispersion since the C interact with its periodic image; we consider the resulting densities at the gamma point as a sufficiently good approximation of the atomic orbital, which is then having a constrained occupation = 1 in the following ΔSCF calculation. The resulting PES curves with each of the three bands populated are shown in Fig. S8. The equilibrium in-plane position of the adsorbate is unchanged since it is dictated by the fourfold symmetry of the adsorption site; furthermore the interaction strength is also little affected, giving a largely unchanged PES. We thus rule out this mechanism for driving in-plane motion.

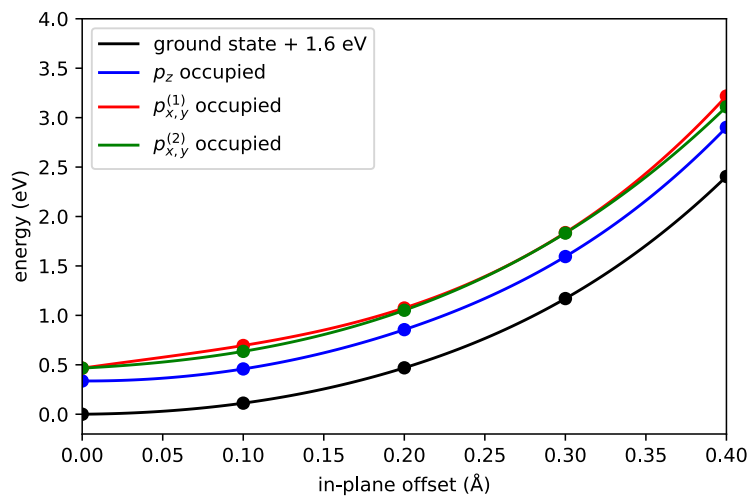

**Fig. S8:** Potential energy curves obtained using the ion resonance model, for the ground state and when populating the three C p-derived orbitals. The ground state curve has been shifted upwards by 1.6 eV to allow easy comparison of the curve shapes.

## REFERENCES

[1]     M. Gajdoš, K. Hummer, G. Kresse, J. Furthmüller, and F. Bechstedt, Phys. Rev. B - Condens. Matter Mater. Phys. **73**, 045112 (2006).

[2]     S. I. Anisimov, B. L. Kapeliovich, and T. L. Perel'man, Zh. Eksp. Teor. Fiz. **66,** 776 (1974) [Sov. Phys.-JETP **39**, 375 (1974)].

[3]     G. K. White, Int. J. Thermophys. **9**, 839 (1988).

[4]     Z. Lin, L. V. Zhigilei, and V. Celli, Phys. Rev. B - Condens. Matter Mater. Phys. **77**, 075133 (2008).

[5]     D. Novko, J. C. Tremblay, M. Alducin, and J. I. Juaristi, Phys. Rev. Lett. **122**, 016806 (2019).

[6] E. Beaurepaire, J.-C. Merle, A. Daunois, and J.-Y. Bigot, Phys. Rev. Lett. **76**, 4250 (1996).

[7] A. P. Caffrey, P. E. Hopkins, J. M. Klopf, and P. M. Norris, Nanoscale Microscale Thermophys. Eng. **9**, 365 (2005).

[8] S.-S. Wellershoff, J. Güdde, J. Hohlfeld, J. G. Müller, and E. Matthias, Proc. SPIE **3343**, 378 (1998).

[9] I. A. Abrikosov, A. V. Ponomareva, P. Steneteg, S. A. Barannikova, and B. Alling, Curr. Opin. Solid State Mater. Sci. **20**, 85 (2016).

[10] B. Koopmans, G. Malinowski, F. Dalla Longa, D. Steiauf, M. Fähnle, T. Roth, M. Cinchetti, and M. Aeschlimann, Nat. Mater. **9**, 259 (2010).

[11] W. You et al., Phys. Rev. Lett. **121**, 077204 (2018).

[12] P. Scheid, G. Malinowski, S. Mangin, and S. Lebègue, Phys. Rev. B **99**, 174415 (2019).

[13] E. Allaria et al., Nat. Photonics **6**, 699 (2012).

[14] T. Katayama et al., J. El. Spect. Rel. Phen. **187**, 9 (2013).

[15] L. Triguero, L. G. M. Pettersson, and H. Ågren, Phys. Rev. B **58**, 8097 (1998).

[16] E. Diesen, G. L. S. Rodrigues, A. C. Luntz, F. Abild-Pedersen, L. G. M. Pettersson, and J. Voss, AIP Adv. **10**, 115014 (2020).

[17] H. Öström et al., Science **347**, 978 (2015).

[18] A. Nilsson et al., Chem. Phys. Lett. **675**, 145 (2017).

[19] H. Öberg et al., Surf. Sci. **640**, 80 (2015).

[20] E. Diesen et al., Phys. Rev. Lett. **127**, 016802 (2021).

[21] P. Giannozzi et al., J. Phys. Condens. Matter **21**, 395502 (2009).

[22] P. Giannozzi et al., J. Phys. Condens. Matter **29**, 465901 (2017).

[23] B. Hammer, L. B. Hansen, and J. K. Nørskov, Phys. Rev. B **59**, 7413 (1999).

[24] S. Mallikarjun Sharada, R. K. B. Karlsson, Y. Maimaiti, J. Voss, and T. Bligaard, Phys. Rev. B **100**, 035439 (2019).

[25] D. Vanderbilt, Phys. Rev. B **41**, 7892 (1990).

[26] M. Taillefumier, D. Cabaret, A.-M. Flank, and F. Mauri, Phys. Rev. B **66**, 195107 (2002).

[27] C. Gougoussis, M. Calandra, A. P. Seitsonen, and F. Mauri, Phys. Rev. B **80**, 075102 (2009).

[28] O. Buñau and M. Calandra, Phys. Rev. B **87**, 205105 (2013).

[29] M. Leetmaa, M. P. Ljungberg, A. Lyubartsev, A. Nilsson, and L. G. M. Pettersson, J. El. Spect. Rel. Phen. **177**, 135 (2010).

[30] L. Triguero, L. G. M. Pettersson, and H. Ågren, J. Phys. Chem. A **102**, 10599 (1998).

[31] A. Föhlisch, J. Hasselström, P. Bennich, N. Wassdahl, O. Karis, A. Nilsson, L. Triguero, M. Nyberg, and L. Pettersson, Phys. Rev. B - Condens. Matter Mater. Phys. **61**, 16229 (2000).

[32] A. Nilsson and N. Mårtensson, Phys. Rev. Lett. **63**, 1483 (1989).

[33] A. Nilsson and L. G. M. Pettersson, Surf. Sci. Rep. **55**, 49 (2004).

[34] G. Kresse and J. Furthmüller, Phys. Rev. B **54**, 11169 (1996).

[35] G. Kresse and J. Furthmüller, Comput. Mater. Sci. **6**, 15 (1996).

[36] G. Kresse and D. Joubert, Phys. Rev. B - Condens. Matter Mater. Phys. **59**, 1758 (1999).

[37] P. E. Blöchl, O. Jepsen, and O. K. Andersen, Phys. Rev. B **49**, 16223 (1994).

[38] J. J. Mortensen, L. B. Hansen, and K. W. Jacobsen, Phys. Rev. B **71**, 035109 (2005).

[39] J. Enkovaara et al., J. Phys. Condens. Matter **22**, 253202 (2010).

[40] K.-i. Inoue, K. Watanabe, and Y. Matsumoto, J. Chem. Phys. **137**, 024704 (2012).

[41] K.-i. Inoue, K. Watanabe, T. Sugimoto, Y. Matsumoto, and T. Yasuike, Phys. Rev. Lett. **117**, 186101 (2016).

[42] See e.g. S. Flügge, *Practical Quantum Mechanics* (Springer-Verlag Berlin Heidelberg, 1971), p. 107-110.

[43] S. Baroni, S. de Gironcoli, A. Dal Corso, and P. Giannozzi, Rev. Mod. Phys. **73**, 515 (2001).

[44] M. Calandra and F. Mauri, Phys. Rev. B **71**, 064501 (2005).

[45] M. Wierzbowska, S. de Gironcoli, and P. Giannozzi, ArXiv Prepr. Cond-Mat/0504077 (2005).

[46] D. Menzel and R. Gomer, J. Chem. Phys. **41**, 3311 (1964).

[47] P. A. Redhead, Can. J. Phys. **42**, 886 (1964).

[48] J. A. Misewich, T. F. Heinz, and D. M. Newns, Phys. Rev. Lett. **68**, 3737 (1992).

[49] J. Gavnholt, T. Olsen, M. Engelund, and J. Schiøtz, Phys. Rev. B **78**, 075441 (2008).